\documentclass[conference]{IEEEtran}
\IEEEoverridecommandlockouts
\usepackage{cite}
\usepackage{amsmath,amssymb,amsfonts}
\usepackage{algorithmic}
\usepackage{graphicx}
\usepackage{textcomp}
\usepackage{xcolor}
\usepackage{glossaries}
\usepackage[binary-units=true]{siunitx}
\usepackage[caption=false]{subfig}

\graphicspath{images/}

\newacronym{DTI}{DTI}{Data Transmission Interval}
\newacronym[plural=BTIs]{BTI}{BTI}{Beacon Transmission Interval}
\newacronym{A-BFT}{A-BFT}{Association Beamforming Training}
\newacronym{ATI}{ATI}{Announcement Transmission Interval}
\newacronym{VR}{VR}{Virtual Reality}
\newacronym[plural=STAs]{STA}{STA}{Station}
\newacronym{CBAP}{CBAP}{Contention-Based Access Period}
\newacronym{SP}{SP}{Service Period}
\newacronym[plural=dynSPs, longplural=dynamic Service Periods]{dynSP}{dynSP}{dynamic Service Period}
\newacronym{PS}{PS}{Pseudo-Static}
\newacronym{NPS}{NPS}{Nonpseudo-Static}
\newacronym{BHI}{BHI}{Beacon Header Interval}
\newacronym{MEC}{MEC}{Mobile Edge Cloud}
\newacronym{EDCA}{EDCA}{Enhanced Distributed Channel Access}
\newacronym{DCF}{DCF}{Distributed Coordination Function}
\newacronym{DIFS}{DIFS}{Distributed Interframe Space}
\newacronym{AIFS}{AIFS}{Arbitration Interframe Space}
\newacronym[plural=ACs,longplural=Access Categories]{AC}{AC}{Access Category}
\newacronym{TXOP}{TXOP}{Transmit Opportunity}
\newacronym{SIFS}{SIFS}{Short Interframe Space}
\newacronym[plural=APs]{AP}{AP}{Access Point}
\newacronym{NAV}{NAV}{Network Allocation Vector}
\newacronym[plural=BIs,longplural=Beacon Intervals]{BI}{BI}{Beacon Interval}
\newacronym[plural=BFs,longplural=Beacon Frames]{BF}{BF}{Beacon Frame}
\newacronym{SPR}{SPR}{Service Period Request}
\newacronym[plural=HMDs, longplural=Head-Mounted Displays]{HMD}{HMD}{Head-Mounted Display}
\newacronym{mmWave}{mmWave}{millimeter wave}
\newacronym{RD}{RD}{Reverse Direction}
\newacronym{MIMO}{MIMO}{Multiple-Input and Multiple-Output}
\newacronym[plural=VFs,longplural=Video Frames]{VF}{VF}{Video Frame}
\newacronym[plural=A-MPDUs, longplural=Aggregated MAC Protocol Data Units]{A-MPDU}{A-MPDU}{Aggregated MAC Protocol Data Unit}

\newacronym{MSDU}{MSDU}{MAC Service Data Unit}
\newacronym{SC}{SC}{Single Carrier}
\newacronym{MCS}{MCS}{Modulation and Coding Scheme}
\newacronym{CTS}{CTS}{Clear To Send}
\newacronym{COTS}{COTS}{Commercial Off-The-Shelf}
\newacronym{CDF}{CDF}{Cumulative Distribution Function}
\newacronym{MTP}{MTP}{Motion-To-Photon}
\def\BibTeX{{\rm B\kern-.05em{\sc i\kern-.025em b}\kern-.08em
    T\kern-.1667em\lower.7ex\hbox{E}\kern-.125emX}}

\begin{document}

\IEEEpubid{
    \begin{minipage}{\textwidth}\ \\\\\\\\\\\\\\ © 2020 IEEE. Personal use of this material is permitted. Permission from IEEE must be obtained for all other uses, in any current or future media, including reprinting/republishing this material for advertising or promotional purposes, creating new collective works, for resale or redistribution to servers or lists, or reuse of any copyrighted component of this work in other works.
      \end{minipage}    } 

\title{Towards Ultra-Low-Latency mmWave Wi-Fi for Multi-User Interactive Virtual Reality\\
}

\author{\IEEEauthorblockN{Jakob Struye, Filip Lemic and Jeroen Famaey}
\IEEEauthorblockA{IDLab - Department of Computer Science\\
University of Antwerp - imec, Antwerp, Belgium\\
Email: \{jakob.struye,filip.lemic,jeroen.famaey\}@uantwerpen.be}
}

\maketitle

\begin{abstract}
The need for cables with high-fidelity Virtual Reality (VR) headsets remains a stumbling block on the path towards interactive multi-user VR. Due to strict latency constraints, designing fully wireless headsets is challenging, with the few commercially available solutions being expensive. These solutions use proprietary millimeter wave (mmWave) communications technologies, as extremely high frequencies are needed to meet the throughput and latency requirements of VR applications. In this work, we investigate whether such a system could be built using specification-compliant IEEE 802.11ad hardware, which would significantly reduce the cost of wireless mmWave VR solutions. We present a theoretical framework to calculate attainable live VR video bitrates for different IEEE 802.11ad channel access methods, using 1 or more head-mounted displays connected to a single Access Point (AP). Using the ns-3 simulator, we validate our theoretical framework, and demonstrate that a properly configured IEEE 802.11ad AP can support at least 8 headsets receiving a 4K video stream for each eye, with transmission latency under 1 millisecond.
\end{abstract}

\section{Introduction}
The interest in \gls{VR} \glspl{HMD} has steadily increased since the field's revitalisation following the announcement of the Oculus Rift. Originally intended as a peripheral for video games, its applications have since broadened to various fields, including healthcare~\cite{applications1}, military and flight training~\cite{applications2}, tourism~\cite{applications3}, and many more. Over the past 5 years, manufacturers including Oculus, HTC, Sony and Valve have all released well-received \glspl{HMD}. However, some widespread restrictions on the format remain. For one, most \glspl{HMD} are wired solutions, tethered to a stationary device responsible for content generation. This restricts users' mobility, reduces immersiveness and represents a tripping hazard. %
The obvious solution is to transmit content wirelessly. The only prominent \gls{HMD} manufacturer currently offering this is HTC, through a wireless add-on for its popular Vive \gls{HMD}, increasing the total cost of the device by half. The add-on communicates in the \SI{60}{\giga\hertz} frequency range using a proprietary protocol developed by Intel. \\%
Another major obstacle, magnified by these wireless solutions, is the \gls{MTP} latency. This type of latency represents the time between the user performing a motion, and the result of this action becoming visible on the \gls{HMD}. Depending on the user, \gls{MTP} latency becomes noticeable between $7$ and \SI{20}{\milli\second}~\cite{towardVR1,towardVR2,vredge}. Apart from network transmission time, \gls{MTP} latency also includes the time needed to sense inputs, computing and processing overheads, and the display's latency. Depending on the hardware used, this leaves between 1 and \SI{5}{\ms} for one-way video transmission. This restriction makes \gls{mmWave} solutions, comprising the \SI{30}{} to \SI{300}{\giga\hertz} frequency range, appealing for these applications, as their inherently high data rates imply that \glspl{VF} can be transmitted faster and, therefore, with lower latency.\\%
In this work, we investigate the applicability of the \gls{mmWave}-based IEEE 802.11ad standard in this domain, for one or more co-located \glspl{HMD}. Specifically, the protocol offers multiple channel access methods, either contention-based or taking a time division approach. We analyse the feasibility of supporting live \gls{VR} with each approach. In live \gls{VR}, content is generated in real-time, dependent on user actions, meaning buffering cannot aid in achieving latency requirements. Current research on \gls{mmWave}'s low-latency capabilities is mostly focused on 5G, not taking any IEEE 802.11ad-specifics into account~\cite{related1, related2, related3, related4}. Works related to IEEE 802.11ad usually focus on only one channel access method, with little to no consideration for the latency of data delivery~\cite{relatedAD1,relatedAD2, relatedAD3}. Furthermore, even latency-focused works on \gls{VR} over IEEE 802.11ad do not take the choice of channel access method and its impact on latency into consideration~\cite{relatedVR1, relatedVR2, relatedVR3, relatedVR4}. In this work, we analyse the attainable video bitrate, and, as an effect, image quality, given a certain upper latency limit and refresh rate, for each of the channel access methods supported by IEEE 802.11ad. We do this because using a standardised protocol, and consequently \gls{COTS} components, is expected to lead to significantly cheaper devices. %
The main goal of this work is to assess whether IEEE 802.11ad is a viable candidate for supporting live \gls{VR} applications, by determining the highest image quality it can support for one or more \glspl{HMD}. In addition, this work forms a basis for future analysis of IEEE 802.11ay in this domain. This standard, which is still a work in progress at the time of writing, is expected to enhance IEEE 802.11ad, reusing and extending its channel access methods~\cite{tdd}. IEEE 802.11ay promises an increase in attainable bitrate by roughly a factor 4, through channel bonding and \gls{MIMO}.\\
The remainder of this paper is structured as follows. Section~\ref{sec:bi} covers IEEE 802.11ad's general structure, and Section~\ref{sec:adVR} analyses its implications for low-latency traffic. In Section~\ref{sec:theory}, we present our theoretical performance analysis, which we validate through simulation in Section~\ref{sec:val}. Finally, Section~\ref{sec:conclusions} concludes this work.

 \section{The IEEE 802.11ad Beacon Interval}\label{sec:bi}
\begin{figure}[!t]
    \centering
    \includegraphics[width=0.48\textwidth]{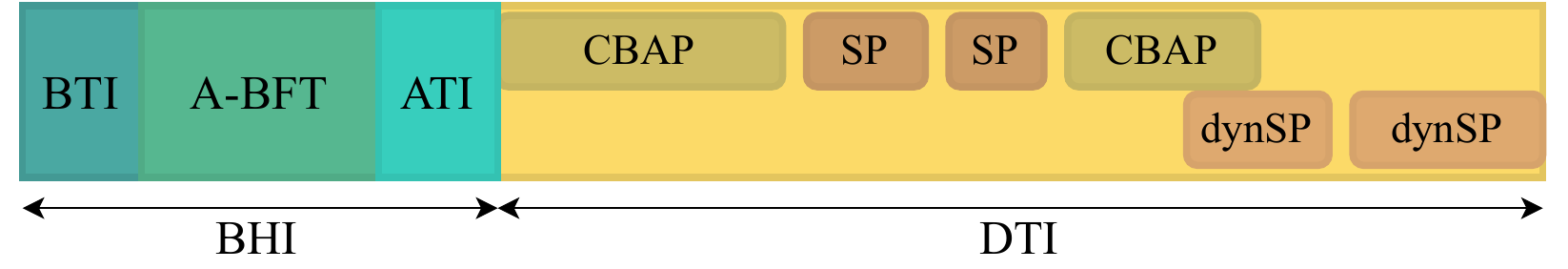}
    \caption{Beacon Interval}
    \label{fig:bi}
\end{figure}

 The IEEE 802.11ad standard divides time into \glspl{BI}~\cite{standard}. A \gls{BI} may take up to \SI{1024.0}{\milli\second}, although \SI{102.4}{\milli\second} is most commonly chosen~\cite{adGeneral1}. The \gls{BI} structure, illustrated in \figurename~\ref{fig:bi}, is divided into two parts: (1) the \gls{BHI}, used for control traffic including association, beamforming and synchronisation, and (2) the \gls{DTI}, where \glspl{STA} may transmit data according to some channel access method. This section covers the internals of these intervals, focusing on their implications in terms of latency.
 \subsection{Beacon Header Interval}
 Compared to similar intervals in other Wi-Fi standards, the \gls{BHI} is rather long and complex. This is largely due to high path loss experienced in the \gls{mmWave} range. Due to legal power emission limits and energy usage concerns, robust \gls{mmWave} links can only be achieved by focusing transmit power in a directional beam, meaning omnidirectional transmission is not feasible. All reachable directions from a \gls{STA} are subdivided into pre-defined \textit{sectors}, and reaching all directions requires sequential transmissions for all sectors.\\
 At the start of the \gls{BHI}, in the \gls{BTI}, the \gls{AP} may transmit \glspl{BF}, informing any STA of its existence, its capabilities, and the specific structure of the remainder of the \gls{BI}. \glspl{BF} use the lowest \gls{MCS}, lengthening transmission. Next, in the \gls{A-BFT} phase, STAs may associate to the AP, and exchange frames with the AP in the beamforming process, in which the optimal sector is selected.
The A-BFT phase is divided into several slots, of which STAs pick one at random in a contention-based approach. Finally, in the \gls{ATI}, the AP can exchange management information with already associated STAs through a unicast, higher-MCS request-response mechanism, which is considerably more spectrally efficient than sending \glspl{BF}~\cite{adGeneral1}. 

\subsection{Data Transmission Interval}
The transmission of actual data (e.g., video content) occurs during the \gls{DTI}. Channel access can be organised with a contention-based approach, using time division with a predefined schedule, or through polling. \glspl{BF} contain an Extended Schedule, which indicates how the following \gls{DTI} is organised. It contains a number of non-overlapping allocations, each assigned one method of channel access. Each allocation can be further subdivided into periods, with each period being equally spaced and equally sized, and periods of different allocations possibly being interleaved.
\subsubsection{Contention-Based Access Period}
The \gls{CBAP} is the simplest type of channel access in IEEE 802.11ad. During a \gls{CBAP}, the well-known \gls{EDCA} algorithm is applied. All incoming data traffic is assigned to one of four \glspl{AC}, each with their own queue, according to latency requirements. Once the medium is sensed to be free for one \gls{AIFS} (of \gls{AC}-dependent duration), a countdown is initialised randomly to an integer between 0 and $cw$ (again \gls{AC}-dependent). The station may commence transmission once this countdown, ticking down once per \SI{5}{\us} slot, reaches 0. 
 Once the \gls{STA} acquires the medium this way, it is granted a \gls{TXOP} of pre-defined, \gls{AC}-dependent, length, during which it may continue transmitting frames of the same \gls{AC}, each separated by 1 \gls{SIFS}. 
When the Extended Schedule is empty, the entire \gls{DTI} may be set to one large \gls{CBAP} through the \texttt{CBAP-only} flag in the \gls{BF}.
\subsubsection{Service Period}
The \gls{SP} is a time division approach. For each \gls{SP}, a pair of \glspl{STA} are appointed as sender and receiver. During the \gls{SP}, the sender has exclusive, uninterrupted access to the medium, but may only send to the configured receiver. If the sender determines that it no longer requires the remainder of its \gls{SP}, it may relinquish the remaining time to the receiver or to the \gls{AP}. %
\subsubsection{Dynamic Allocation of Service Periods}\label{dynalloc}
In case of bursty, non-periodic traffic patterns, the \gls{SP} mechanism is far from optimal. It is therefore also possible to create \glspl{SP} dynamically, based on demand, during the \gls{DTI}. %
These \glspl{dynSP} are announced by sending Grant frames, optionally preceded by the \gls{AP} polling \glspl{STA} for grant requests. These Grant frames can be sent during \gls{CBAP} or \gls{SP} allocations, and a \gls{dynSP} may overlap with or exceed the allocation during which it was announced. %
\Glspl{dynSP} too can be truncated.

\section{IEEE 802.11ad for Low Latency Traffic}\label{sec:adVR}
The exact organisation of the \gls{BI} has severe implications on latency-sensitive traffic, such as in live \gls{VR}. Both the \gls{BHI} and \gls{DTI} need to be carefully organised to minimise their impact on the latency of content delivery.
\subsection{Beacon Header Interval Optimisation} 
The length of the BHI sets a lower bound on the attainable worst-case latency in the network, as no data transmission may occur during it. A relatively small 70 byte \gls{BF} already takes upwards of \SI{33}{\us} to transmit per sector~\cite{standard}. Furthermore, a single A-BFT slot of an 8-sector AP takes \SI{173}{\us}. Taking into account interframe spaces and propagation time, a BHI for 8 sectors with the default 8 A-BFT slots takes \SI{1.664}{\ms}, with the optional ATI disabled entirely. This alone prevents the network from achieving sub-ms latencies consistently. Fortunately, there are a number of opportunities to decrease the BHI length. First, the BTI is not mandatory in every BHI, as the standard only requires it being present once every 15 BIs. However, the AP is required to send a \gls{BF} on each sector at least once every 4 BIs. As such, the AP can rotate through sectors between BTIs, ideally dividing the number of \glspl{BF} by 4. Next, the A-BFT is also required only once per 15 BIs, and its number of slots can be as low as 1. Lowering the number of slots only impacts performance when regular beamforming is needed due to \gls{STA} mobility or environment dynamics, which are out of scope in this work. Overall, these two improvements reduce the worst-case BHI duration of an 8-sector \gls{AP} to \SI{249}{\us}, including \SI{10}{\us} of interframe spaces.\\%
This \gls{BHI} configuration has a number of side-effects. First, STAs will, by design, no longer receive a \gls{BF} for every BI. For such BIs without a \gls{BF}, the STA does not know which allocations were assigned within the \gls{DTI}. However, to alleviate this issue, allocations can be marked \gls{PS}. These allocations are assumed to reoccur for 4 BIs, starting from the one its allocation was received in, each time at the same offset from the start of the BI. %
A \gls{DTI}-spanning \gls{CBAP} allocation, indicated through the \texttt{CBAP-only} flag, is also considered to be \gls{PS}. As such, the reduced number of \glspl{BF} has no effect on STAs' ability to participate in data transfer during \gls{PS} allocations, as long as no \glspl{BF} are lost.
\begin{figure}[!t]
    \centering
    \subfloat[\gls{CBAP} overheads\label{fig:cbap_overheads}]{\includegraphics[width=0.45\textwidth]{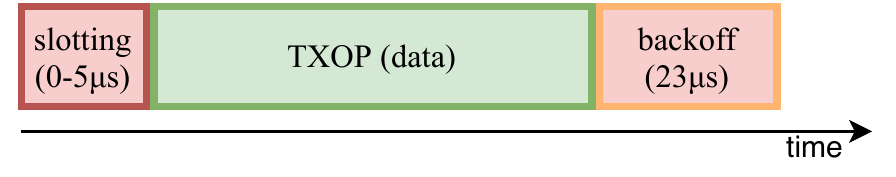}}\\
    \subfloat[\gls{SP} overheads\label{fig:sp_overheads}]{\includegraphics[width=0.45\textwidth]{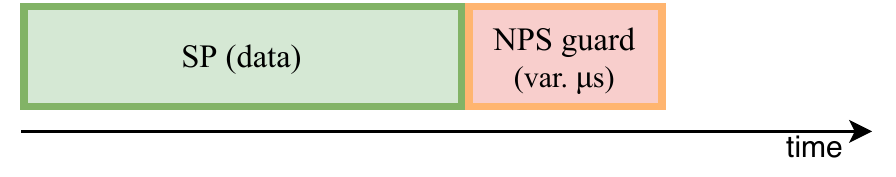}}\\
    \subfloat[\gls{dynSP} overheads\label{fig:dynsp_overheads}]{\includegraphics[width=0.45\textwidth]{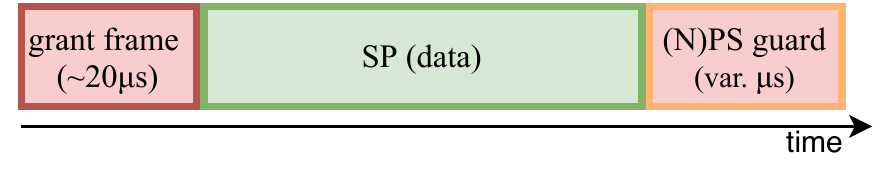}}\\
    \caption{Overheads resulting from the channel access mechanism. Overheads preceding \gls{VF} transmission (shown in red) must occur when data transmission is imminent. Overheads following data transmission (shown in orange) must finish before the following \gls{VF} commences. Guard time length increases with \gls{BI} length. Further analysis of overhead duration is presented in Section \ref{sec:vfblock}.}
    \label{fig:overheads}
\end{figure}
\subsection{Data Transmission Interval Optimisation}
All three types of channel access incur their own set of overheads, summarised in \figurename~\ref{fig:overheads}. An obvious overhead of \gls{CBAP} is the time spent in the channel sensing and the backoff periods before transmission is allowed. 
However, a \gls{STA} may enter its backoff period for an \gls{AC} even if no frames are currently queued for it. Once the backoff timer expires, the system enters a \textit{post-backoff} state~\cite{postBackoff}. If, within this state, a frame arrives in the queue, transmission may begin at the start of the next \SI{5}{\us} backoff slot. With optimal settings and no competing \glspl{STA}, the post-backoff state can be reached after observing the medium for, at most, \SI{23}{\us}. 
Next, by making sure the \gls{TXOP} limit is configured to be sufficient to transmit a full \gls{VF}, only a single \gls{TXOP} is needed for each \gls{VF}. An overdimensioned \gls{TXOP} limit has no negative side effects, as the sender can end the \gls{TXOP} early simply by refraining from sending any more data.\\%
For a scheduled \gls{SP}-based system, no slotting overhead exists. However, tight synchronisation between the content server and \gls{AP} is crucial. The \gls{AP} must be aware of the video streams' characteristics for \gls{SP} scheduling, and \glspl{SP} have to be shifted every \gls{BI} to maintain synchronisation, meaning only \gls{NPS} allocations can be used. With \glspl{dynSP}, Grant frames add latency. The allocations for Grant frame transmission may be \gls{PS} or \gls{NPS}.\\
Another important latency factor is the use of guard times. A guard time must occur between any two subsequent allocations, and ahead of a \gls{CBAP}-only allocation. As each \gls{STA}'s clock may drift from the clock provided in the \glspl{BF}, these guard times are necessary to ensure that adjacent allocations' transmissions do not overlap. The minimum guard time $g_i$, in \SI{}{\micro\second}, between allocations $i$ and $i+1$ is defined as:
\begin{equation}
 g_i = \left\lceil\frac{(A_i C D_i) + (A_{i+1} C D_{i+1})}{10^6} + SIFS + T_p\right\rceil\label{eq:guard},
\end{equation}
 where $A_i$ is 5 for \gls{PS} allocations %
 and 1 otherwise, $C$ is the maximum allowable clock drift, defined as \SI{20}{ppm}, $D_i$ is the time passed since the latest synchronisation (or the \gls{BI} length for \gls{PS} allocations), the \gls{SIFS} is \SI{3}{\micro\second}, and $T_p$ is the air propagation time between two \glspl{STA}, defined as \SI{0.1}{\micro\second}. Guard times for \gls{PS} allocations are significantly longer than for \gls{NPS} allocations, although the exclusive use of \gls{PS} allocations does shorten the \gls{BHI}. In addition, guard times grow as the \gls{BI} length increases. The precise impact is investigated in the following section.

\section{Theoretical Analysis}\label{sec:theory}
In this section, we apply our findings in a \gls{mmWave} multi-user \gls{VR} environment, determining its maximum attainable per-user bitrate.
\subsection{Virtual Reality Setup}
\begin{figure}[!t]
    \centering
    \includegraphics[width=0.48\textwidth]{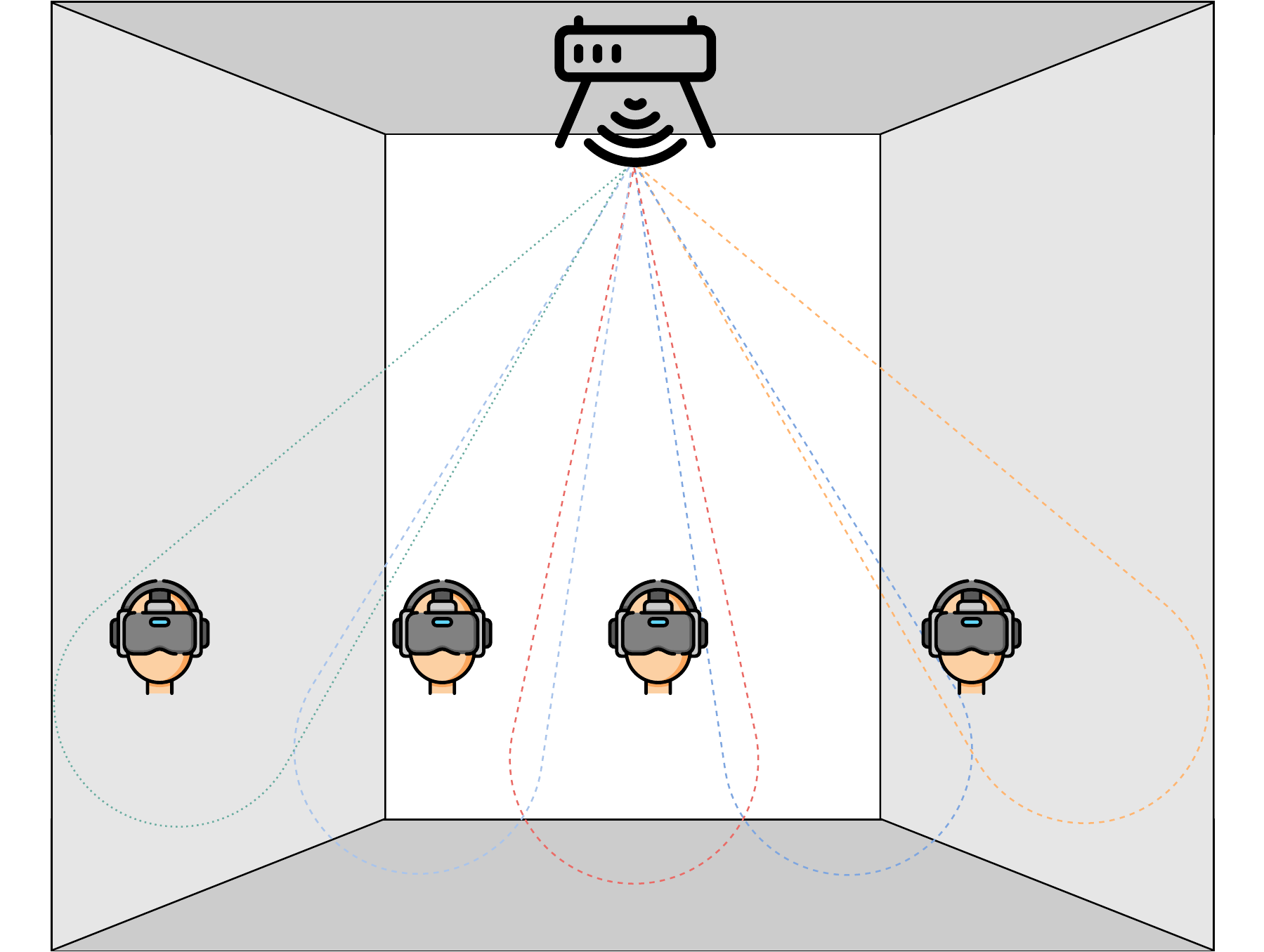}
    \caption{The \gls{VR} room setup. 1 ceiling-mounted \gls{AP} uses 8 beams (5 drawn) to serve up to 8 \glspl{HMD} (4 drawn) on the ground. Some \gls{HMD} movement is allowed, but each user is assumed to stay within one beam's reach.}
    \label{fig:setup}
\end{figure}
We consider an obstacle-free room with one or more \gls{HMD}-wearing users on the ground, and a single ceiling-mounted central \gls{AP}, as shown in \figurename~\ref{fig:setup}. In this initial work, we assume that the \gls{AP} placement guarantees an unobstructed line of sight with each \gls{HMD} and we do not explicitly account for significant user mobility, but note that some user movements can occur. As long as the \gls{HMD} remains within one sector, mobility should not affect connectivity. Current-day IEEE 802.11ad \glspl{AP} are limited to 2-8 relatively wide sectors. Beamforming is assumed to have been performed in advance, and its optimisation is considered out of scope in this work. All devices use the \gls{SC} PHY at the maximum \gls{MCS} 12. %
The \gls{AP} is directly connected to a \textit{content server} (possibly a Mobile Edge Cloud) responsible for \gls{VF} generation and processing for all users. \glspl{VF} are generated in real-time, at a fixed framerate, and immediately transmitted to the users one-by-one. Network-wise, the video content is streamed over UDP, chosen for its low overhead. %
At the MAC layer, the \gls{AP} aggregates data using \glspl{A-MPDU}, as this again lowers overhead. One such \gls{A-MPDU} can fit at most 32 data units, each containing \SI{7884}{\byte} of application data (plus \SI{66}{\byte} of headers up to the transport layer). %
We only consider downstream traffic, but note that our findings are easily extended to also consider some upstream traffic, such as viewing direction, voice, and user inputs. %
\subsection{Abstractions}
Given a system with $n$ \glspl{HMD} running at a refresh rate $r$ and a maximum allowed \gls{VF} transmission latency $l_{max}$, our goal is to find the maximum attainable video bitrate $b$ that will not violate the \gls{VF} transmission latency. To compare latency under different channel access methods, we abstract all types of latencies that may delay \gls{VF} delivery into one of three classes. First is the \textit{interBI latency}, which only occurs once per \gls{BI}, at its start. This relatively rare but long latency block comprises the \gls{BHI}, any guard time preceding the first allocation in the \gls{BI}, and any latency before the \gls{AP} can access the medium during this allocation, induced by the channel access method. Next is the regular \textit{inter\gls{VF} latency}, occurring between any two subsequent \gls{VF} transmissions (unless overridden by interBI latency) and immediately following the previous transmission. This includes guard times between allocations and, again, any latency before the \gls{AP} can access the medium during the allocation, induced by the channel access method. Finally, \textit{access latency} occurs between a \gls{VF}'s arrival at the \gls{AP} and the start of its transmission. This comprises any latency induced by the channel access method, occurring regardless of the observed medium state before the \gls{VF} arrived. This may include overheads due to slotting, and control overhead that must occur just before data transmission. Note that any channel access method-agnostic overheads, such as PHY/MAC headers and RTS/CTS overheads, are accounted for in Section~\ref{sec:bitrate}.\\ We divide time into \textit{\gls{VF} intervals} of length $1/r$, such that, for each \gls{HMD}, exactly one \gls{VF} is generated per \gls{VF} interval. The \gls{VF} interval consists of $n$ latency blocks (at most 1 interBI, and $n-1$ or $n$ inter\gls{VF}) and $n$ equally-sized \gls{VF} blocks, each available for transmission to one \gls{HMD}. For convenience, we define access latency to be part of the \gls{VF} block. By analysing how much time of the \gls{VF} interval is lost to these types of latencies, the time available for \gls{VF} transmission for each \gls{HMD} can easily be calculated. Note that only the worst case is considered; often the interBI latency will not be present, replacing it with the significantly shorter inter\gls{VF} latency. %
\subsection{Coordination Levels}
\begin{figure}[!t]
    \centering
    \includegraphics[width=0.45\textwidth]{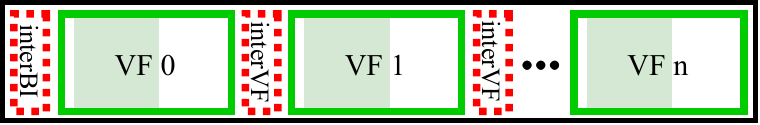}
    \caption{The VF interval with BI coordination, with \gls{VF} blocks (solid, green) and latency blocks (dotted, red). The shaded part of the \gls{VF} blocks, of length $v_{tx}$, can be used for transmission.}
    \label{fig:VF_BI}
  \end{figure}

  \begin{figure}[!t]
    \centering
    \subfloat[The VF interval as intended at content server\label{fig:VF_video_server}]{\includegraphics[width=0.45\textwidth]{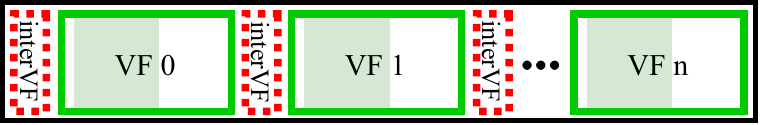}}\\
    \subfloat[The VF interval as executed at \gls{AP}\label{fig:VF_video_AP}]{\includegraphics[width=0.45\textwidth]{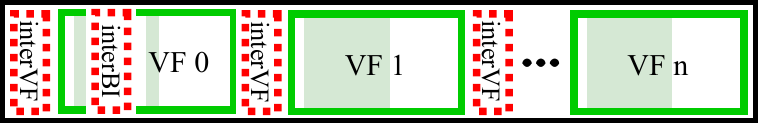}}
    \caption{The VF interval with video coordination}
    \label{fig:VF_video}
\end{figure}
Depending on the exact physical setup and customisability of the \gls{AP}, different levels of coordination may be feasible. We consider two cases: (1) tight coordination between content server and \gls{AP}, with the content server being \textit{\gls{BI}-aware}, and (2) coordination between the different video streams at the \gls{AP}.
\subsubsection{BI coordination}\label{tightcoord}
In this case, the content server is aware of the general IEEE 802.11ad \gls{BI} structure, and carefully schedules \gls{VF} generation to not overlap with any latency blocks. %
Without loss of generality, we assume that each interBI latency block occurs at the start of a \gls{VF} interval. The full \gls{VF} interval is illustrated in Fig.~\ref{fig:VF_BI}. After determining interBI latency $l_{iBI}$ and inter\gls{VF} latency $l_{iVF}$, the maximum length of a \gls{VF} block $v$ is easily calculated:
\begin{equation}
v = \frac{\frac{1}{r} - l_{iBI} - (n-1) l_{iVF}}{n}\label{eq:pb}.
\end{equation}
 Access latency $l_{acc}$ and maximum allowed latency $l_{max}$ however limit how much of the \gls{VF} block may be used for data transmission. We therefore divide $v$ into three parts: an access latency part of length $v_{pre}$, a usable part of length $v_{tx}$, and an unused \textit{end buffer} of length $v_{buf}$. These lengths are calculated as $v_{pre} = l_{acc}$, $v_{tx} = min(v,l_{max}) - l_{acc}$ and $v_{buf} = max(0,v-l_{max})$, such that $v = v_{pre} + v_{tx} + v_{buf}$.

\subsubsection{Video coordination}
\begin{figure}[!t]
    \centering
    \begin{minipage}{1.0\linewidth}
    \subfloat[Greedy scheduling\label{fig:schedule_greedy}]{\includegraphics[width=0.47\textwidth]{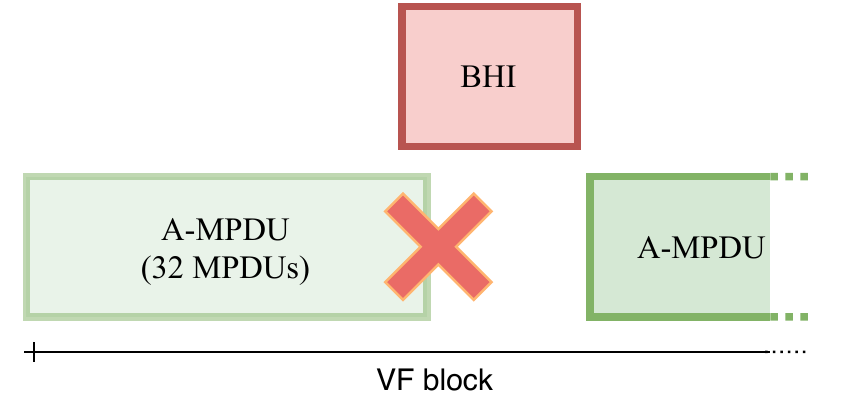}}
    \hfill
    \subfloat[Smart scheduling\label{fig:schedule_smart}]{\includegraphics[width=0.47\textwidth]{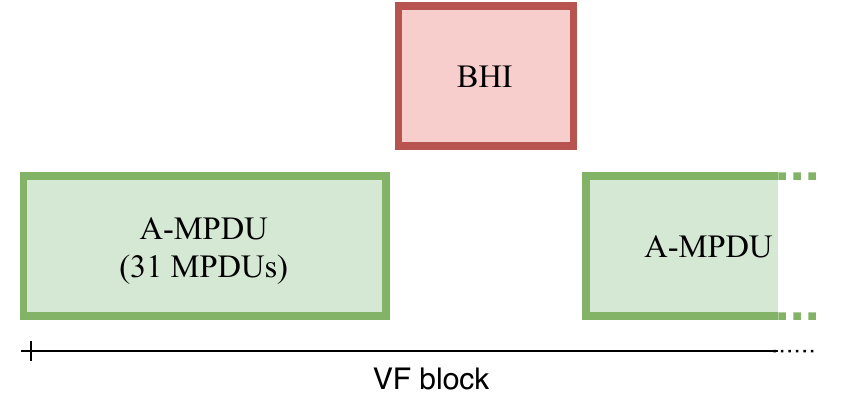}}
    \end{minipage}
    \caption{Usually, an \gls{AP} will fill an \gls{A-MPDU} with as many MPDUs as possible before attempting transmission. In the above case, the transmission time available ahead of the \gls{BHI} was not enough for a full \gls{A-MPDU} (see \protect\subref{fig:schedule_greedy}) but could have accommodated a non-full one (see \protect\subref{fig:schedule_smart}). While such smart scheduling could increase throughput significantly as shown here, we assume no such system is available on the \gls{AP}, as it would be challenging to implement to run in real-time.}
    \label{fig:scheduletypes}
\end{figure}

\begin{figure}[!t]
    \centering
    \subfloat[Maximum throughput with fortunate \gls{BHI} placement\label{fig:shortvf_optimal}]{\includegraphics[width=0.23\textwidth]{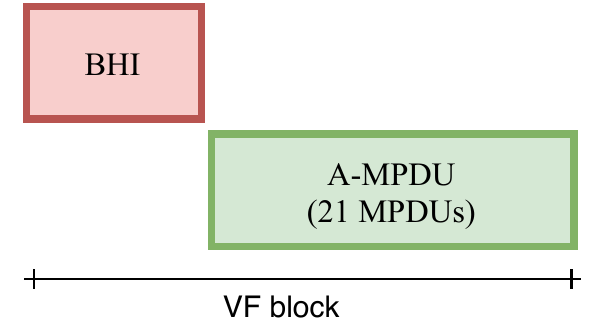}}\\
    \begin{minipage}{1.0\linewidth}
    \subfloat[Bitrate too high, failed tx\label{fig:shortvf_bad}]{\includegraphics[width=0.49\textwidth]{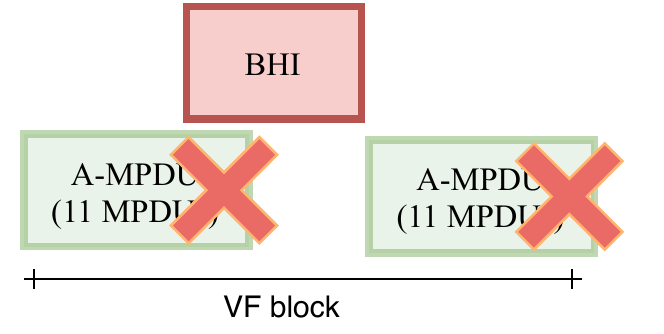}}
    \hfill
    \subfloat[Appropriate bitrate (1)\label{fig:shortvf_good1}]{\includegraphics[width=0.49\textwidth]{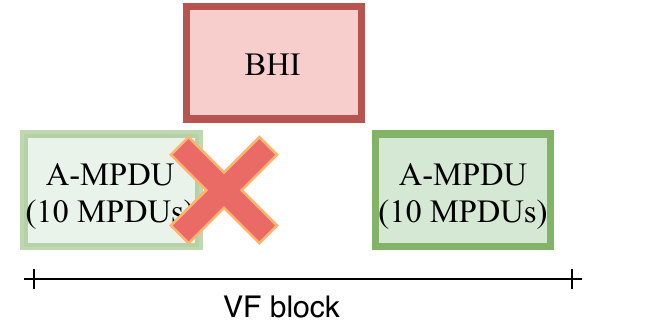}}
    \end{minipage}\\
    \begin{minipage}{1.0\linewidth}
    \subfloat[Appropriate bitrate (2))\label{fig:shortvf_good2}]{\includegraphics[width=0.49\textwidth]{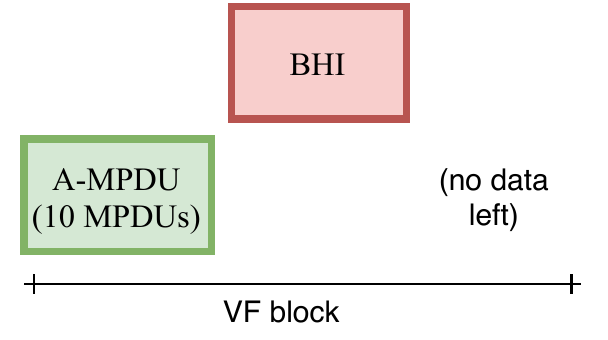}}
    \hfill
    \subfloat[Appropriate bitrate (3)\label{fig:shortvf_good3}]{\includegraphics[width=0.49\textwidth]{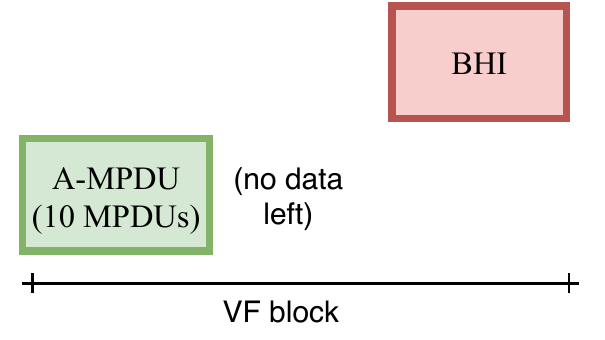}}
    \end{minipage}%
    \caption{In this situation, a full \gls{A-MPDU} can never be sent during a \gls{VF} block reduced by an interBI block (consisting mainly of the \gls{BHI}). \protect\subref{fig:shortvf_optimal} shows that, with fortunate interBI placement, at most 21 MPDUs will fit. With suboptimal interBI placement however, an \gls{A-MPDU} of only 11 MPDUs may fail to transmit, as shown in \protect\subref{fig:shortvf_bad}. At most half of the optimal 21 MPDUs can always be sent sent successfully, regardless of the exact interBI block placement, as illustrated in \protect\subref{fig:shortvf_good1}-\protect\subref{fig:shortvf_good3}. Smart scheduling, as illustrated in \figurename~\ref{fig:scheduletypes}, would alleviate this phenomenon, but is assumed to not be supported by the \gls{AP}.}
    \label{fig:shortvf}
\end{figure}

In the second case, the content server no longer actively attempts to avoid interBI latency blocks. Instead, it simply divides \gls{VF} blocks evenly across the \gls{VF} interval. While the content server still leaves room for the inter\gls{VF} latency block (whose position is decided by the preceding \gls{VF} block), a \gls{VF} block may now overlap with an interBI latency block. As a result, the transmission schedule as intended by the content server, may differ from that actually used at the \gls{AP}. When an interBI latency block is inserted during \gls{VF} transmission, the \gls{AP} may slice the \gls{VF} block in two, such that $v = v_{pre1} + v_{tx1} + v_{pre2} + v_{tx2} + v_{buf}$. %
Fig. \ref{fig:VF_video_server} and \ref{fig:VF_video_AP} show the schedule as intended at the content server, and executed at the \gls{AP}, respectively. 
In the worst case, the interBI latency block is scheduled such that the first part of the \gls{VF} block is just too short to send the first \gls{A-MPDU}. Unless the \gls{AP} can dynamically adapt its maximum \gls{A-MPDU} size given the time remaining in the current allocation (which would be challenging to implement, and therefore unlikely to be supported by \gls{COTS} hardware), $v_{tx1}$ cannot actually be used for data transmission if it is shorter than $t_{aggr}$, the time needed to successfully complete a full \gls{A-MPDU} transmission (calculated in Section \ref{sec:bitrate}). As long as 1 \gls{VF} requires at least 1 full \gls{A-MPDU}, this worst-case $v_{tx1}$ remains unused, and all data transmission only occurs in $v_{tx2}$. If instead a single non-full \gls{A-MPDU} suffices, it could be sent in either $v_{tx1}$ or $v_{tx2}$, whichever is biggest, meaning the worst case occurs when the two are equal. \figurename~\ref{fig:scheduletypes} and \figurename~\ref{fig:shortvf} illustrate these two cases. The actually usable $v_{tx}$ in both cases can be defined as:
\begin{equation}
    v_{tx} = max(\frac{v_{tx1}+v_{tx2}}{2}, v_{tx1}+v_{tx2}-t_{aggr})\label{eq:videocoord1}
\end{equation}
\begin{equation}
    v_{tx1}+v_{tx2} = min(v,l_{max})-l_{iBI} - 2l_{acc}\label{eq:videocoord2}
\end{equation}


\subsection{VF Block Length}\label{sec:vfblock}
We now calculate \gls{VF} block length $v$ for each combination of coordination assumption and channel access method. Recall that \figurename~\ref{fig:VF_BI} and \ref{fig:VF_video} summarise the \gls{VF} interval structure, while \figurename~\ref{fig:overheads} details the \gls{VF} block structure for each channel access method, with access latency in red and interVF latency in orange. For every method with \gls{PS} allocations, each sector sees one \gls{BF} once every four \glspl{BI}, while with \gls{NPS} allocations each sector receives a \gls{BF} every \gls{BI}. The full \gls{BHI} for 8 sectors has a fixed base transmission time of \SI{249}{\us} (\gls{PS} only) or \SI{453}{\us} (incl. \gls{NPS}), increased by \SI{5}{\us} per \gls{BF} for each allocation in the Extended Schedule. The guard time as defined in \eqref{eq:guard} occurs between every pair of adjacent allocations and before a \gls{CBAP}-only allocation. Channel sensing and the backoff period between two \glspl{TXOP} amount to at most \SI{23}{\micro\second}. %
Finally, the access latency comprises waiting at most \SI{5}{\micro\second} for the next slot start in \gls{CBAP} allocations, the \SI{19.8}{\us} transmission of a Grant frame for \glspl{dynSP}, and is zero in all other cases. Table \ref{table:latencies} shows all latencies, assuming an 8 sector \gls{AP} and a \gls{BI} length equal to the \gls{VF} interval length. The two components of the \gls{BHI} duration are listed separately, between brackets. Given the interBI and interVF latency, we calculate $v$ for 1, 2, 4 and 8 \glspl{HMD}, with refresh rate \SI{120}{\Hz}, shown in Table \ref{table:blockLengths}. As long as $v > l_{max}$, the value of $l_{max}$ has a direct, significant influence on the attainable video bitrate, meaning any hardware or software improvements lowering other aspects of \gls{MTP} latency can indirectly increase this bitrate. %

\begin{table}[!t]
\renewcommand{\arraystretch}{1.1}
\caption{Latency Block Lengths for an 8-sector \gls{AP} with $n$ \glspl{HMD}, in \SI{}{\us}}
\centering
\begin{tabular}{cccc}
\hline
\multicolumn{1}{c}{}& \multicolumn{1}{c}{\textbf{interBI}} & \multicolumn{1}{c}{\textbf{interVF}} & \multicolumn{1}{c}{\textbf{access}} \\

\hline
\textbf{CBAP-only} & $(253+2\cdot0)$ + 5 & 23 &5\\
\textbf{PS CBAP} & $(253+2\cdot5)$ & 23 & 5\\
\textbf{NPS CBAP} & $(453+8\cdot5)$ & 23 & 5\\
\textbf{NPS SP} & $(453+n\cdot8\cdot5)$ & 4 & 0\\
\textbf{PS dynSP} & $(253+2\cdot5)$ & 5 & 19.8\\
\textbf{NPS dynSP} & $(453+8\cdot5)$ & 4 & 19.8\\
\hline
\end{tabular}
\label{table:latencies}
\end{table}

\begin{table}[!t]
    \renewcommand{\arraystretch}{1.1}
    \caption{\gls{VF} Block Length $v$ at \SI{120}{\Hz}, in \SI{}{\ms}}
    \centering
    \begin{tabular}{ccccc}
    \hline
    \multicolumn{1}{c}{}& \multicolumn{1}{c}{\textbf{1 HMD}} & \multicolumn{1}{c}{\textbf{2 HMDs}} & \multicolumn{1}{c}{\textbf{4 HMDs}} & \multicolumn{1}{c}{\textbf{8 HMDs}} \\
    
    \hline
    \textbf{CBAP-only} & 8.079 & 4.026 & 1.999 & 0.985 \\
    \textbf{PS CBAP} & 8.074 & 4.023 & 1.998 & 0.985 \\
    \textbf{NPS CBAP} & 7.840 & 3.906 & 1.939 & 0.956 \\
    \textbf{NPS SP} & 7.840 & 3.898 & 1.927 & 0.942 \\
    \textbf{PS dynSP} & 8.074 & 4.035 & 2.015 & 1.005 \\
    \textbf{NPS dynSP} & 7.840 & 3.918 & 1.957 & 0.977 \\
    \hline
    \end{tabular}
    \label{table:blockLengths}
\end{table}

\subsection{Attainable Bitrate}\label{sec:bitrate}
Given the latency block lengths for a configuration, we can calculate $v_{tx}$, the time available for data transmission, and convert this to a video bitrate. The \gls{AP} sends a number of \glspl{A-MPDU}, each requiring only 1 PHY header, and acknowledged with a single Block ACK. The PHY sends 1 chip per \SI{0.57}{\ns}, translating to \SI{4620}{Mbps} at \gls{MCS} 12~\cite{standard}. %
As such, transmission of 1 \gls{A-MPDU} consists of, in order: 1 PHY preamble + header ($7552+1024$ chips), 32 MPDUs (each \SI{7950}{\byte}), 1 \gls{SIFS}, 1 PHY preamble + header, 1 Block ACK (\SI{32}{\byte}), 1 \gls{SIFS}. %
The duration of one \gls{A-MPDU} $t_{aggr}$ then becomes:
\begin{equation*}
    t_{aggr} = 2 t_{PHY} + t_{BA} + 2 SIFS +  32 t_{MPDU}
\end{equation*}
where $t_{PHY}$ is the preamble and PHY header overhead, and $t_{BA}$ and $t_{MPDU}$ are the MAC-level transmission times of a Block ACK and an MPDU, respectively. The number of full \glspl{A-MPDU} $a$ that can be sent in 1 $v_{tx}$ then becomes
\begin{equation*}
    a = \left\lfloor\frac{v_{tx} + 2 SIFS + t_{PHY} + t_{BA}}{t_{aggr}}\right\rfloor
\end{equation*}
and finally one more non-full \gls{A-MPDU} of $b$ MPDUs can be sent, if $b>0$:
\begin{equation*}
    b = \left\lfloor\frac{v_{tx} - a t_{aggr} - t_{PHY}} {t_{MPDU}}\right\rfloor
\end{equation*}
The total attainable size for one \gls{VF} then becomes $(32a+b)$\SI{7884}{\byte}. This is easily translated to video bitrate, given the refresh rate. Table \ref{table:throughputsCombined} shows the attainable bitrate for 1 and 8 \glspl{HMD}, given the latency block lengths in Table \ref{table:latencies}, with refresh rate $r$=\SI{120}{\Hz}. The impact of adding more \glspl{HMD} is limited with coordination; the additional \glspl{HMD} mainly reduce $v_{buf}$. Overall, \gls{PS} approaches are more viable, as their \gls{BHI} is significantly shorter, while their higher guard times are barely noticeable. Guard times scale with \gls{BI} length, which we chose to be only $1/r$. Conveniently, the top-performing \gls{PS} \gls{CBAP} and \gls{CBAP}-only approaches are also the simplest to implement, and therefore %
most likely to be supported by \gls{COTS} hardware.\\
Finally, note that RTS/CTS could easily be taken into consideration by subtracting its overhead from $v_{tx}$. Similarly, upstream traffic could easily be sent in the end buffer, as the \gls{STA} can be granted channel access in the \gls{TXOP}/\gls{SP} through the Reverse Direction protocol~\cite{standard}. If the end buffer does not suffice, $v_{tx}$ could again be reduced.

\begin{table}[!t]
    \renewcommand{\arraystretch}{1.1}
    \caption{Throughputs at \SI{120}{\Hz}, in Mbit/s}
    \centering
    \begin{tabular}{ccccccc}
    \hline
    \multicolumn{1}{c}{}& \multicolumn{3}{c}{\textbf{BI coordination}} & \multicolumn{3}{c}{\textbf{Video coordination}} \\
    \multicolumn{1}{c}{}& \multicolumn{2}{c}{\textbf{1 HMD}} & \multicolumn{1}{c}{\textbf{8 HMDs}} & \multicolumn{2}{c}{\textbf{1 HMD}} & \multicolumn{1}{c}{\textbf{8 HMDs}} \\
    \multicolumn{1}{c}{}& \multicolumn{1}{c}{\textbf{1ms}} & \multicolumn{1}{c}{\textbf{5ms}} & \multicolumn{1}{c}{\textbf{1ms}} & \multicolumn{1}{c}{\textbf{1ms}} & \multicolumn{1}{c}{\textbf{5ms}} & \multicolumn{1}{c}{\textbf{1ms}} \\
    
    \hline
    \textbf{CBAP-only} & 505 & 2541 & 498 & 188 & 2187 & 188  \\
    \textbf{PS CBAP} & 505 & 2541 & 498 & 188 & 2180 & 180\\
    \textbf{NPS CBAP} & 505 & 2541 & 484 & 123 & 2064 & 115\\
    \textbf{NPS SP} & 505 & 2548 & 476 & 115 & 2050 & 29\\
    \textbf{PS dynSP} & 498 & 2541 & 498 & 180 & 2165 & 180\\
    \textbf{NPS dynSP} & 498 & 2541 & 484 & 130 & 2072 & 123 \\
    \hline
    \end{tabular}
    \label{table:throughputsCombined}
\end{table}

\section {Validation}\label{sec:val}
\begin{figure}[!t]
    \centering
    \includegraphics[width=0.45\textwidth]{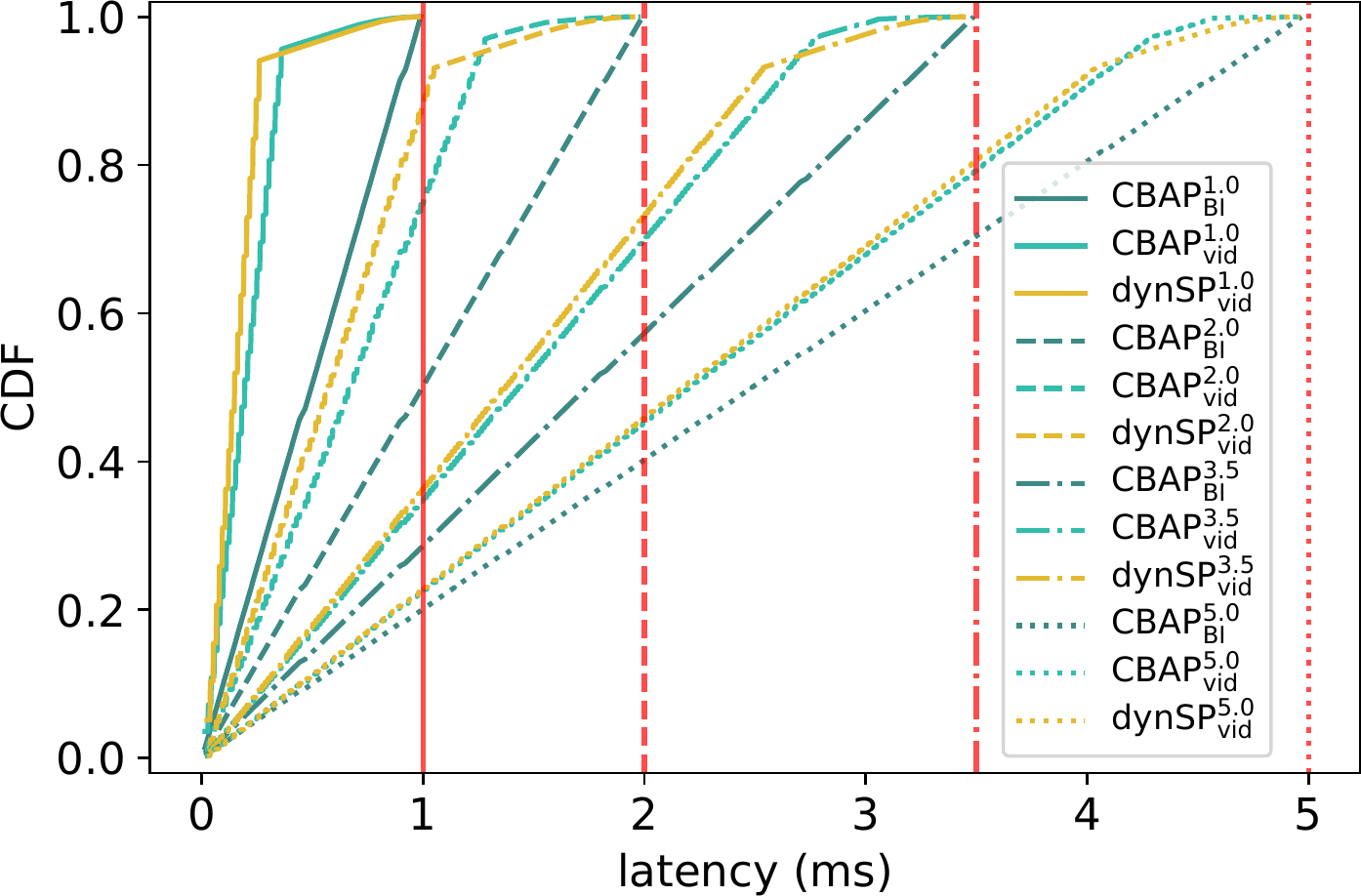}

    \caption{Packet latency CDF as simulated in ns-3 for different $l_{max}$ values (superscript), indicated with red lines.}
    \label{fig:cdf}
  \end{figure}

We now validate our theoretical results using the IEEE 802.11ad module~\cite{ns3ad1,ns3ad2} of the ns-3 simulator~\cite{ns3}. We evaluate three combinations of channel access method and coordination level, repeating the experiments for four different $l_{max}$ values: $1.0$, $2.0$, $3.5$ and \SI{5.0}{\ms}. The used bitrates are partially found in Table \ref{table:throughputsCombined}, the others can be calculated with the formulas presented. We measure the latency of each \gls{VF}-carrying packet (between the end of \gls{VF} generation and delivery at the \gls{HMD}), and show the \gls{CDF} for all experiments in Fig.~\ref{fig:cdf}. %
We first validate the BI-coordinated \gls{CBAP}-only approach. We implement the coordination by slightly increasing the refresh rate, such that the \gls{BI} length is a multiple of the \gls{VF} interval length, and shift the maximum attainable bitrates accordingly. For this single-\gls{HMD} experiment, labelled CBAP\textsubscript{BI}, latency approaches $l_{max}$ in each case, but never exceeds it. The highest latencies observed are \SI{0.990}{\ms}, \SI{1.992}{\ms}, \SI{3.488}{\ms} and \SI{4.969}{\ms}.\\ %
Second, we validate the video coordination approach. %
As in the previous case, all packets in this experiment, labelled CBAP\textsubscript{vid}, arrive on time as intended, with highest latencies \SI{0.982}{\ms}, \SI{1.984}{\ms}, \SI{3.478}{\ms} and \SI{4.984}{\ms}. Notice that the long tail of the \gls{CDF} is indicative of the \gls{BHI}, configured to occur every \SI{10.24}{\ms}, occasionally overlapping with \gls{VF} blocks. %
\def\SPSBbottom#1#2{\rlap{\textsuperscript{#1}}\textsubscript{#2}}
Third, we repeat this experiment with \glspl{dynSP} using \gls{PS} allocations, labelled dynSP\textsubscript{vid}, again reaching the same conclusion, with highest latencies \SI{0.986}{\ms}, \SI{1.959}{\ms}, \SI{3.453}{\ms} and \SI{4.961}{\ms}. This experiment exhibits an even longer tail, as its \gls{BHI} is significantly longer. As a final experiment, we validate our analysis for multi-\gls{HMD} setups by repeating the CBAP\SPSBbottom{1.0}{vid} case for 8 \glspl{HMD}, which, as expected, shows no difference in latency compared to the single-\gls{HMD} case.
\section {Conclusions and Future Work}\label{sec:conclusions}
In this work, we presented the first comparison of IEEE 802.11ad's different channel access methods with regards to latency-sensitive live \gls{VR} traffic. Specifically, we provided a theoretical framework for deriving the maximum attainable bitrates within given latency bounds for each access method. Through this framework, we demonstrated the severe impact of beacon transmission on the attainable video bitrate. In addition, we showed that the use of \glsfirst{PS} allocations, as well as tight coordination between content server and \gls{AP}, can significantly improve said bitrates.\\
Entry-level \glspl{HMD}, with two 2K displays, require a throughput of \SI{100}{Mbit}~\cite{vredge}, which we have demonstrated to be attainable at a transmission latency of only \SI{1}{\ms}, with any channel access method and for at least 8 \glspl{HMD}, assuming the frames of the different video streams are properly interleaved. If the content server is BI-aware, at least 8 advanced \glspl{HMD}, featuring 4K displays and each requiring \SI{400}{Mbit}, can be supported with a transmission latency of \SI{1}{\ms}. Thus, our work suggests IEEE 802.11ad as a viable candidate in supporting live \gls{VR} applications. Future ultimate \gls{VR}~\cite{ultimateVR}, featuring 8K displays and requiring \SI{1.5}{Gbps} can only be supported at a transmission latency of \SI{5}{\ms}. Lowering this to \SI{1}{\ms} will require the additional throughput offered by IEEE 802.11ay. In our future work, we will explore the limits of live \gls{VR} over IEEE 802.11ay, and characterise the effects of interference and \gls{HMD} mobility on achievable bitrates and latency guarantees.

\section*{Acknowledgment}
The work of Jakob Struye was supported by the Research Foundation - Flanders (FWO): PhD Fellowship 1SB0719N. The work of Filip Lemic was supported by the EU Marie Skłodowska- Curie Actions Individual Fellowships (MSCA-IF) project Scalable Localization-enabled In-body Terahertz Nanonetwork (SCaLeITN), grant nr. 893760. In addition, this work received support from the University of Antwerp's University Research Fund (BOF). The authors thank Hany Assasa for support on the ns-3 IEEE 802.11ad module.

\end{document}